\documentclass[fleqn,twoside]{article}
\usepackage{espcrc2}

\usepackage{graphicx}

\newcommand{\AmS}{{\protect\the\textfont2
  A\kern-.1667em\lower.5ex\hbox{M}\kern-.125emS}}

\newcommand{\dslash}[2]{{{#1}\hspace{-5pt}{/}}_{#2}}
\newcommand{\Bsl}{{B\hspace{-5pt}{/}}}
\newcommand{\ndop}{{\mathcal O}_{R/L;L}^{\Bsl}}
\newcommand{\msbar}{\overline{\mbox{MS}}}

\def\simge{
    \mathrel{\rlap{\raise 0.511ex
        \hbox{$>$}}{\lower 0.511ex \hbox{$\sim$}}}}
\def\simle{
    \mathrel{\rlap{\raise 0.511ex
        \hbox{$<$}}{\lower 0.511ex \hbox{$\sim$}}}}

\title{
\vspace{-3.6cm}
\begin{flushright}
{\normalsize
hep-lat/0409114\\
WUB-04-11\\
}
\end{flushright}
\vspace{2cm}
Nucleon Decay Matrix Elements with $N_f=0$ and $2$ Domain-Wall Quarks}

\author{Yasumichi Aoki\address{Physics Department, University of Wuppertal, 
Gaussstr 20, 42119 Wuppertal, Germany. }
[RBC collaboration]
\thanks{We thank RIKEN, BNL and the U.S.\ DOE for their support
to this project.}}

\begin{document}

\begin{abstract}
 The nucleon decay matrix elements of three-quark operators
 are calculated with domain-wall fermions. Operators are
 renormalized non-perturbatively to match the $\msbar$ (NDR)
 scheme at NLO. Quenched simulation
 studies involve both direct measurement of the matrix elements and
 the chiral Lagrangian parameters, $\alpha$ and $\beta$. We also
 report on the dynamical quark effects on these parameters.
\end{abstract}

\maketitle

\section{INTRODUCTION}

Nucleon decay \cite{ndecay},
once observed, is the hallmark of 
the physics beyond the standard model.
On-going deep mine experiments, though yet to observe an event, are
pushing up the lower bound of the lifetime of the
nucleon \cite{Shiozawa:2003it2}, excluding  GUT models 
which allow nucleons decay more frequently.

The dominant decay mode of the nucleon 
is to a pseudoscalar meson and a lepton, where low energy hadronic interactions
are important. The factorization technique, commonly used
for the hadron decays, leads to the decay amplitude written in terms
of the Wilson coefficients and the low-energy matrix elements of the
dimension-six 
operator consisting of three quarks and one lepton. Various QCD model
calculations had estimated 
the hadronic part of the matrix element with results varying over
an order of magnitude (see a compilation in \cite{Brodsky:1984st}).
The amplitude must be squared in the width or lifetime,
meaning two orders of magnitude different lifetime from different
estimations.
The initial lattice calculations show smaller but, still large
deviation (see a compilation in \cite{Aoki:1999tw}).

Recent lattice calculations 
concentrate on how to remove the systematic uncertainties,
especially the discretization error,
by either taking the continuum limit \cite{Tsutsui:2004qc} of the Wilson
fermion \cite{Aoki:1999tw} or employing the domain-wall fermion 
(DWF)
\cite{Aoki:2002ji,Aoki:2003vz}. All these calculations, however, have
been done within the quenched approximation.

We report here the updated results of the nucleon decay matrix elements
with DWF with quenched approximation as well as 
the estimate of the low energy parameters with $N_f=2$ dynamical 
DWF. Detailed calculations will be given in the
forthcoming publication \cite{AokiRBC:3000ya}.

\section{QUENCHED SIMULATION}

The parameters of our quenched simulation are $\beta=0.87$ for the
gauge coupling of the DBW2 gauge action, domain-wall height $M_5=1.8$,
fifth dimension $L_s=12$, the lattice volume $16^3\times 32$ where
latter is the temporal direction. The inverse lattice spacing
$a^{-1}\simeq 1.3$ GeV is obtained by the $m_\rho=0.77$ GeV input at the chiral
limit, which gives the physical spatial lattice size as $L\simeq 2.4$ fm.
While we only have one lattice spacing, the scaling violation is
expected to be small for DWF as the other
hadronic observables have shown \cite{Aoki:2002vt}.

The renormalization of the operator is done in combination of the
non-perturbative renormalization (NPR)
with
the continuum perturbation
theory in  NLO.  The NPR scheme \cite{Martinelli:1995ty} is employed for
the lattice operator to match 
the MOM scheme. The one-loop matching factor that we have calculated 
converts it to the $\msbar$ in NDR scheme. Then, using the two-loop anomalous
dimension \cite{Nihei:1995tx}, the operators are run to $\mu=2$ GeV.

We demonstrated the NPR work well last year \cite{Aoki:2003vz}
reusing the data for the quark bilinear renormalization at $L_s=16$.
We have performed NPR calculation at $L_s=12$,
where the matrix 
elements are obtained as well, to get the proper renormalization
factor and to confirm the basic properties such as the absence of the
mixing. 51 configurations are used for NPR.

We also showed the lattice value of the hadronic matrix elements
in the previous reports \cite{Aoki:2002ji,Aoki:2003vz}.
While the number of independent gauge configurations analyzed in this 
study is unchanged (100 configurations), we make use of discrete
symmetry transformation  
properties which relate the different Green functions to be averaged.
In this way the effective statistics have been increased .
Further we have changed the analyses details to get more robust results.
All these have changed the results slightly and helped shrinking the error.

\begin{figure}[t]
 \begin{center}
  \includegraphics[width=7.5cm]{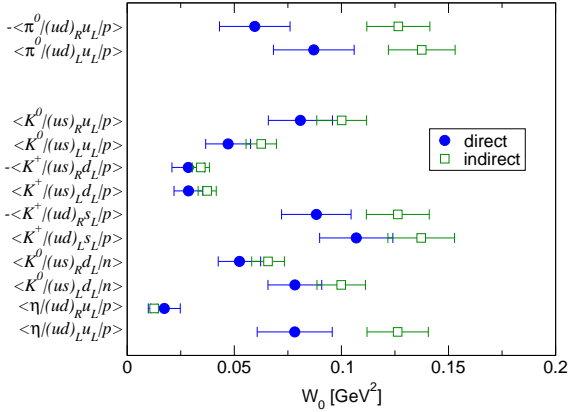}
 \end{center}
 \vspace{-32pt}
 \caption{Summary of $W_0$ (Eq.~\ref{eq:W0}) with direct and indirect
 method in quenched calculation. Operators are renormalized at $\mu=2$ GeV.}
 \label{fig:W0}
 \vspace{-20pt}
\end{figure}

In Fig.~\ref{fig:W0} we show the results of the relevant form factor
$W_0$, which, for example in the $p\to \pi^{0/+}$ decay, defined through
\cite{Aoki:1999tw},
\begin{equation}
 \langle \pi;\vec{p} | \ndop | p;\vec{k}\rangle = 
  P_{L} [ W_0 - i \dslash{q}{} W_q] u_p,
  \label{eq:W0}
\end{equation}
where $\ndop =\epsilon^{ijk}(u^{iT} CP_{R/L}d^j)P_{L}u^k$,
$q_\mu$ is the momentum transfer from proton ($k_\mu$) to pion ($p_\mu$),
$q_\mu=k_\mu-p_\mu$, $u_p$ is the proton spinor.
The results both with direct and indirect calculations are shown.
The former involves various two- and three-point functions with many
different parameter values to allow extrapolation to the physical kinematics
point, thus is expensive. The latter is obtained with help of tree-level
chiral perturbation theory  
\cite{Claudson:1982gh,Aoki:1999tw} where the low energy
parameters $\alpha$ and $\beta$, defined as
\begin{equation}
 \alpha P_L u_p =
  \langle 0 | {\mathcal O}_{R;L}^{\dslash{B}{}} | p\rangle,\; 
 \beta P_L u_p =
 \langle 0 | {\mathcal O}_{L;L}^{\dslash{B}{}} | p\rangle,
\end{equation}
are calculated on the lattice. The other low energy parameters are
taken from the experiment \cite{Aoki:2003vz}.
The indirect method uses only a few two point functions, thus less
computational effort is required than the direct method.
The SU$(2)$ flavor symmetry of $u$ and $d$ quarks relates the
different matrix elements, e.g.,\
\begin{equation}
 \langle \pi^+| \ndop | p\rangle= \sqrt{2}\langle \pi^0| \ndop | p\rangle.
  \label{eq:su2}
\end{equation}
Every other possible matrix element is identical to one of those in
Eq.~\ref{eq:su2} or Fig.~\ref{fig:W0} up to sign factor.

The final $\pi$ state matrix elements has apparent deviation between
the direct and indirect calculations.  This may simply mean the limit
of the tree-level  $\chi$PT, as the pion has large momentum.
Indeed in the soft pion limit ($p_\mu\to 0$)
the expected relation $(W_0-iq_4W_q)^{R;L} = \alpha/\sqrt{2}f_\pi$ holds
numerically  within the error.
We note that the final $K$ state matrix elements are estimated better
than those for the $\pi$ by the indirect method.

The values of individual matrix elements are different from those 
obtained for $a^{-1}\simeq 2.3$ GeV with Wilson fermion 
\cite{Aoki:1999tw},  
while ratios of the matrix elements are similar.
We consider our DWF results are closer to the continuum limit.
Indeed it is the case for the low energy parameters.

The results of $\alpha$ and $\beta$ are summarized in the Table
\ref{tab:alpha}. They have a different relative sign and the relation
$\alpha+\beta=0$ approximately holds with $1.3\sigma$.
Recent quenched calculation results \cite{Tsutsui:2004qc} with Wilson fermion 
in the continuum limit are also shown. There the operators are
renormalized by an improved
lattice perturbation theory. We remark that the DWF results
at coarser lattice are consistent with them.

\section{DYNAMICAL QUARK EFFECTS}

To investigate the quenching error we calculate the $\alpha$ and
$\beta$ parameters on the $N_f=2$ dynamical DWF configurations
\cite{IzubuchiRBC:3000tp}.  The simulation parameters are the
same as quenched, except for gauge coupling $\beta=0.8$, which
gives $a^{-1}\simeq1.7$ GeV, $L\simeq 1.9$ fm by $m_\rho$ in the chiral
limit. We have three dynamical quark masses ($m_{dyn}$) spanning 
$m_s/2 \simle m_{dyn} \simle m_s$, where $m_s$ is the strange
quark mass. The spatial lattice size divided by the pion Compton wave
length  is $L m_\pi=4.7$ at the smallest.
We expect the systematic error from the finite volume
effects is subdominant given the large statistical error of the
matrix elements. The operator renormalization is done in the
same way as quenched calculation. 
The number of configurations used is $\sim 100$ for matrix elements
or $\sim 40$ for NPR for each dynamical mass.

\begin{figure}[t]
 \begin{center}
  \includegraphics[width=7cm]{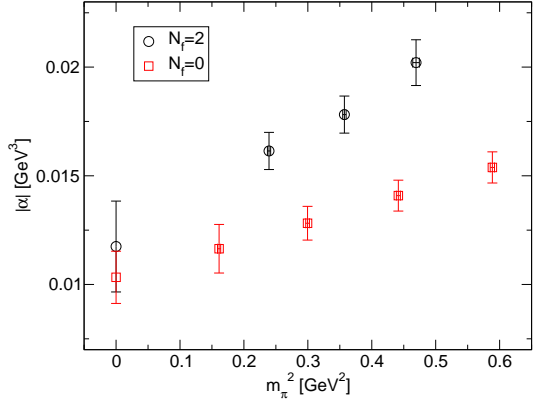}
 \end{center}
 \vspace{-32pt}
 \caption{Pion mass dependence of $\alpha$ in both dynamical ($N_f=2$) and
 quenched ($N_f=0$) calculations.}
 \label{fig:alpha}
 \vspace{-24pt}
\end{figure}

The measurements are done at the valence masses equal to the 
dynamical masses. The pion mass dependence of $|\alpha|$ for
both quenched and dynamical calculations are shown
in Fig.~\ref{fig:alpha}. The dynamical result has stronger $m_\pi$
dependence than quenched.
After rather long extrapolation to the chiral limit
with linear function of quark mass we obtain the $\alpha$ and $\beta$
parameters 
as shown in the Table \ref{tab:alpha}. The relation $\alpha+\beta=0$
holds within the error. The values are consistent with the
quenched ones. We note that these estimates from lattice QCD lie in the
middle between the smallest ($0.003$ GeV$^3$ \cite{Donoghue:1982jm}) and
largest ($0.03$ GeV$^3$ \cite{Brodsky:1984st}) estimates of QCD model
calculations. 
\begin{table}[htb]
 \caption{Summary of the low energy parameter of nucleon decay at the
 renormalization scale $\mu=2$ GeV.
 Quoted errors for DWF are statistical only.}
 \label{tab:alpha}
 \begin{center}
 \begin{tabular}{|c||c|cc|}
  \hline
  Fermion & Wilson\cite{Tsutsui:2004qc} & \multicolumn{2}{c|}{DWF}\\
  \hline
  $N_f$ & 0  & 0 & 2 \\
  $a$ [fm] & 0 & 0.15 & 0.12 \\
  \mbox{\hspace{-4pt}}$|\alpha|$ [GeV$^3$]\mbox{\hspace{-4pt}} & \mbox{\hspace{-4pt}}0.0090(09)($^{+5}_{-19}$)\mbox{\hspace{-4pt}} & 0.010(1) & 0.012(2) \\
  \mbox{\hspace{-4pt}}$|\beta|$  [GeV$^3$]\mbox{\hspace{-4pt}} & \mbox{\hspace{-4pt}}0.0096(09)($^{+6}_{-20}$)\mbox{\hspace{-4pt}} & 0.011(1) & 0.012(2) \\
  \hline
 \end{tabular}
 \end{center}
 \vspace{-24pt}
\end{table}

\section{OUTLOOK}

The present dynamical DWF simulation with quark masses $m_{dyn}\simge m_s/2$
has not revealed the unquenching effect on the $\alpha$ and $\beta$
parameters in the chiral limit with $20\%$ statistical error.
To reduce the error to $10\%$ level, much lighter
dynamical quark region needs to be explored. 
As to this direction, the direct calculation of the matrix elements becomes
more important. Furthermore,
test of scaling to the continuum limit will also be needed.

\bibliography{proc}
\bibliographystyle{h-elsevier}

\end{document}